# Spin Induced Optical Conductivity in the Spin Liquid Candidate Herbertsmithite

D. V. Pilon[1], C. H. Lui[1], T. H. Han[1], D. B. Shrekenhamer[2], A. J. Frenzel[1,3], W. J. Padilla[2], Y. S. Lee[1], N. Gedik[1]*

[1] Department of Physics, Massachusetts Institute of Technology, Cambridge, Massachusetts 02139, USA
[2] Department of Physics, Boston College, Chestnut Hill, Massachusetts 02467, USA
[3] Department of Physics, Harvard University, Cambridge, Massachusetts 02138, USA

*Corresponding author (email: gedik@mit.edu)

A quantum spin liquid (QSL) is a state of matter in which magnetic spins interact strongly, but quantum fluctuations inhibit long-range magnetic order even at zero temperature. A QSL has been predicted to have a host of exotic properties, including fractionalized excitations[1-3] and long-range quantum entanglement[3]. Despite the numerous theoretical studies, experimental realization of a QSL has proved to be challenging due to the lack of candidate materials. The triangular organic salts EtMe$_3$Sb[Pd(dmit)$_2$]$_2$[4,5] and $\kappa$-(BEDT-TTF)$_2$Cu$_2$(CN)$_3$[6-9], and kagome ZnCu$_3$(OH)$_6$Cl$_2$ (Herbertsmithite)[9-17] have recently emerged as promising candidates of exhibiting a QSL state, but the nature of their ground states is still elusive. Here we studied a large-area high-quality single crystal of Herbertsmithite by means of time-domain terahertz (THz) spectroscopy. We observed in the low-frequency (0.6-2.2 THz) optical conductivity evidence for the nature of the spin system. In particular, the in-plane absorption spectrum exhibits a unique frequency dependence that can be described by a power-law with an exponent of approximately 1.4, in sharp contrast with the $\omega^4$ dependence expected for an ordered Mott insulator[18]. The absorption is also found to increase as the temperature decreases, a behavior unexpected for conventional insulators. Such features are consistent with recent theory[18] based on the interactions between the charge and spin degrees of freedom in a QSL system.

The concept of a quantum spin liquid (QSL) was first conceived of by Anderson in 1973[19] and later suggested to be a possible explanation for high temperature superconductivity in the cuprates[20,21]. The proposed ground state of this novel phase, the so-called Resonating Valence Bond (RVB) state, hosts exotic excitations through spin-charge separation, giving rise to chargeless spin ½ spinons, in contrast to the conventional spin-wave excitations (magnons with spin 1) in ordered Mott insulators[1-3,22]. While QSL's have only been a theoretical construct for decades, recent experiments have provided compelling evidence that the long-sought QSL system can be realized in the kagome-lattice antiferromagnet Herbertsmithite [10-14,17,23,24] (Figure 1a). For instance, no magnetic order is observed in this material down to T = 50 mK in thermodynamic measurements[11], and a continuum of spin excitations is exhibited in the inelastic neutron scattering[17].

Despite these experimental works, there are still a number of open questions on the nature of the ground state in Herbertsmithite, specifically the gauge group of the spin liquid state and the character of the low-energy spin excitations. A recent computational study has proposed the existence of a Z$_2$ spin liquid with a sizable spin gap[25], but thermodynamic and inelastic neutron scattering results have shown no sign of a spin gap down to 0.1 meV[11,17], suggesting the existence of a gapless U(1) spin liquid state. Optical studies have proved difficult due to the chargeless nature of the spinon excitations and the relatively low energy scale in the spin system. Recent theoretical studies have, however, proposed that spin-charge interactions through a gauge field in a U(1) Dirac spin liquid can give rise to a contribution to the low-frequency optical conductivity[18,26]. In particular, a power-law dependence of the conductivity ($\sigma$) on the photon frequency ($\omega$), i.e. $\sigma \sim \omega^\beta$, with an exponent $\beta = 2$, is expected at frequencies far below the charge gap. Direct measurement of such power-law optical conductivity is therefore a great step toward elucidating the nature of the QSL ground state and the structure of the low-energy excitations in Herbertsmithite.

In this letter, we report a direct observation of the spin-induced low-frequency optical conductivity in Herbertsmithite. This experiment is made possible for the first time by the recent successful growth of large-area single crystals of Herbertsmithite[27]. By using terahertz (THz) time-domain spectroscopy, we have measured the optical conductivity of Herbertsmithite as a function of temperature and magnetic field in the spectral range of 0.6 to 2.2 THz. Remarkably, the in-plane conductivity $\sigma_{ab}(\omega)$, which is associated with the spin liquid state in the kagome (*ab*) planes of Herbersmithite, is found to depend on frequency as $\sigma_{ab}(\omega) \sim \omega^\beta$ with $\beta \approx 1.4$, a result compatible with the theoretical predictions[18]. The observed power-law conductivity also exhibits the opposite temperature dependence expected for an insulator, and is absent in the out-of-plane direction, as expected for a two-dimensional gapless spin liquid state. Our results are consistent with the predicted low-frequency absorption arising from an emergent gauge field in a gapless U(1) Dirac spin liquid.

In our experiment, we investigated a large single-crystal sample of Herbertsmithite with dimensions 3 x 6 x 0.8 mm. The sample was characterized by neutron diffraction, anomalous x-ray diffraction, and thermodynamic measurements, with results compatible with those of powder samples[27]. As shown in Figure 1a, Herbertsmithite has a layered structure, with planes of spin ½ copper atoms in a kagome pattern separated by the nonmagnetic zinc atoms. This material exhibits strong in-plane (*ab*) antiferromagnetic interactions with Curie-Weiss temperature $\Theta_{CW}$ = -300 K, a charge gap of ~2 eV and a spinon gap of less than 0.1 meV, as well as an exchange energy $J \approx 17$ meV (200 K) as described by a Heisenberg model, with negligible out-of-plane (*c*) interactions[11]. Despite the strong interactions in this material, geometric frustration prevents the formation of any magnetic order down to T = 50 mK[11]. In our experiment, we measured the THz in-plane conductivity $\sigma_{ab}(\omega)$, which is associated with the spin liquid state in the crystal. THz pulses were focused onto the sample at normal incidence, with polarization along the kagome (*ab*) planes. From the transmission of THz radiation, we extracted the frequency-dependent complex optical conductivity, taking into account the sample geometry.

Figure 2a displays the in-plane optical conductivity spectra $\sigma_{ab}(\omega)$ in the frequency range of 0.6 - 2.2 THz at temperatures from 4 K to 150 K. The conductivity spectra can be described by two components. The higher-frequency component, which is significant for frequencies > 1.4 THz, can be attributed to the phonon absorption with a resonance at ~3 THz (see Supplementary Information). Here we focus on the lower-frequency component, which dominates the absorption at frequencies < 1.4 THz. This component can be described by a power law with a small exponent $\sigma_{ab}(\omega) \sim \omega^{\beta}$, where $\beta \approx 1.4$. (We note that, due to the limited frequency range in our measurement, an exponent $\beta$ between 1 and 2 is still compatible with the data.) Such an absorption behavior is distinct from that expected in ordered Mott insulators, which typically exhibit $\omega^4$ frequency-dependent conductivity at low frequencies arising from the spin-wave excitations[18]. In the following discussion, we will provide evidence that the $\omega^\beta$ absorption arises from the spin excitations in Herbertsmithite.

First, we observed a noticeable enhancement of the $\omega^\beta$ absorption component as the temperature decreased from 150 K to 4 K (Figure 2b). The increase of absorption at lower temperature, we remark, is a rather anomalous phenomenon for insulating materials, where light absorption far below the band gap typically decreases at low temperature due to the freezing of phonons and doped charges. The observed unusual temperature dependence immediately indicates that the underlying absorption mechanism must be of exotic origin. Indeed, the temperature dependence of $\sigma_{ab}(\omega)$ is reminiscent of that of metals, suggesting that $\sigma_{ab}(\omega)$ is associated with a gapless or nearly-gapless spin system in Herbertsmithite. Our results are consistent with a similar phenomenon in the Raman scattering of Herbertsmithite, where a continuum of Raman signal due to spinon excitations is found to increase with decreasing temperature at T < 50 K[15]. From the lower bound of our measured frequency range, we estimate that the spin gap in Herbertsmithite, if it exists, should not be larger than 0.6 THz (~2 meV). This value is consistent with the upper bound of the spin gap (~0.1 meV) estimated by other experimental studies [11,17].

Second, we found that the $\omega^\beta$ absorption component disappears in the direction perpendicular to the kagome planes. We have measured the out-of-plane conductivity $\sigma_c(\omega)$ along the *c*-axis of the Herbertsmithite crystal at different temperatures (see Supplementary Information). The $\sigma_c(\omega)$ spectra resemble the tail of a phonon resonance at >3 THz (Figure 3). In particular, $\sigma_c(\omega)$ at T = 4 K is found to drop rapidly to zero at frequencies lower than 1 THz, in contrast to the $\omega^\beta$ behavior observed in the in-plane conductivity $\sigma_{ab}(\omega)$. In addition, the magnitude of $\sigma_c(\omega)$ is found to drop as the temperature decreases (Figures 2b&3), a typical behavior of phonon absorption. The absence of the $\omega^\beta$ absorption component in the out-of-plane direction indicates that it is adherent to the in-plane properties of the sample. Due to the quasi-two-dimensional nature of the spin system in Herbertsmithite, where the spin excitations are confined to move only in the kagome planes, our result strongly suggests that the $\omega^\beta$ absorption found only within the planes is associated with the spin degree of freedom in Herbertsmithite.

Third, the $\omega^\beta$ absorption is unaffected by the presence of a strong magnetic field. We have measured the in-plane conductivity $\sigma_{ab}(\omega)$ in the spectral range of 0.6 - 2.2 THz under magnetic fields from zero to 7 T at T = 6 K (Figure 4). We did not observe any systematic changes of the absorption spectrum with the magnetic field. This rules out any magnetic impurities as the source of the observed THz absorption (see Supplementary Information). Indeed, the result is consistent with the expected behavior of the spinon excitations in a QSL, which generally exhibits a field-independent energy spectrum except in extreme conditions (such as in B > 12 T or T < 0.5 K, where a phase transition may occur.)[28,29].

Finally, the observed frequency dependence and magnitude of $\sigma_{ab}$ agree well with those predicted for the spin induced absorption in a QSL. In particular, recent theory showed that in a gapless U(1) Dirac spin liquid a power-law optical conductivity $\sigma \sim \omega^\beta$ with $\beta = 2$ can arise from the spin-charge interactions through an internal gauge field, and the calculated absorption magnitude is compatible with our experimental data[18]. Similar power-law absorption is also suggested for a gapped $Z_2$ spin liquid[23] due to modulation of the Dzyaloshinkii-Moriya (DM) interaction, but the predicted absorption magnitude is three orders of magnitude too small to match our data. Our results therefore favor the existence of a U(1) spin liquid state with an emergent gauge field in Herbertsmithite.

In conclusion, we have observed a power-law component $\omega^\beta$ with $\beta = 1\sim2$ in the low-frequency in-plane optical conductivity of spin liquid candidate Herbertsmithite. Detailed analysis supports that the absorption arises from the spin excitations. Our results agree with theoretical predictions based on the spin-charge coupling through an emergent gauge field in a gapless Dirac spin liquid, and put an upper bound of 2 meV on the size of the spin gap. This discovery has profound implications on research on quantum spin liquids and hints at the existence of a gauge field in a spin liquid. More generally, our research demonstrates terahertz spectroscopy to be an effective probe to study quantum spin liquids. Conductivity measurements have the potential to extend to the GHz and MHz frequency range via electronic methods, which, combined with sub-Kelvin cooling, may allow us to probe the extremely low-energy excitations in the spin liquid and provide a definite answer on the nature of the ground state in Herbertsmithite.

**Methods**

**Terahertz time-domain spectroscopy** Our experiment was conducted using a mode-locked Ti:Sapphire amplified laser system that generated pulses with 800-nm central wavelength, 100-fs pulse duration and 5-kHz repetition rate. The THz radiation was generated via optical rectification in a <110>-oriented ZnTe crystal, focused onto the sample using the off-axis parabolic mirrors, and subsequently detected by free space electro-optic sampling in a second ZnTe crystal. The polarization of the THz radiation was controlled using a wire-grid polarizer. The sample was mounted in a He$^4$ cryostat for the temperature dependent measurements. A superconducting magnetic cryostat was used for the magnetic-field-dependent measurements.

**Acknowledgement** We thank P. A. Lee, A. Potter, T. Senthil and C. Wang for discussions. This work was supported by grants from the US Department of Energy (DOE), Office of Science, Office of Basic Energy Sciences (BES) (grants DE-FG02-08ER46521 and DE-SC0006423 to N.G., grant DE-FG02-07ER46134 to Y.S.L).

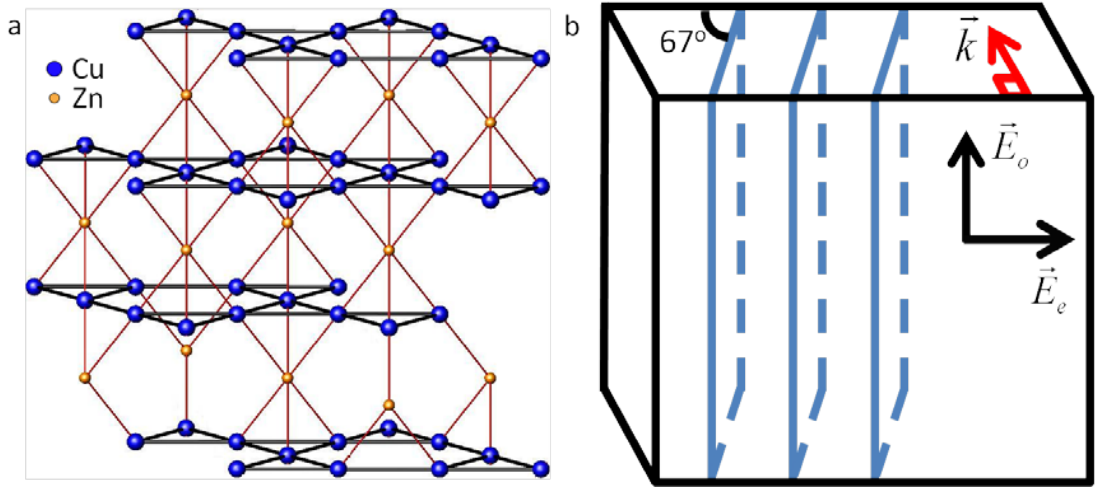

**Figure 1 | The structure and orientation of the Herbertsmithite sample.** **a.** The lattice structure of Herbertsmithite. The crystal is composed of layers of spin ½ copper atoms (blue) arranged in the kagome pattern in the *ab*-plane, separated by nonmagnetic zinc atoms (orange) along the perpendicular *c*-axis. The chlorine, oxygen, and hydrogen atoms in Herbertsmithite are neglected here for simplicity. **b.** The orientation of the sample used in our experiment. The *ab*-planes are aligned vertically, making an angle of 67° with the sample surface. $E_o$, $E_e$ and $k$ denote, respectively, the ordinary and extraordinary optical axes of the crystal and the wave vector of the incident terahertz beam.

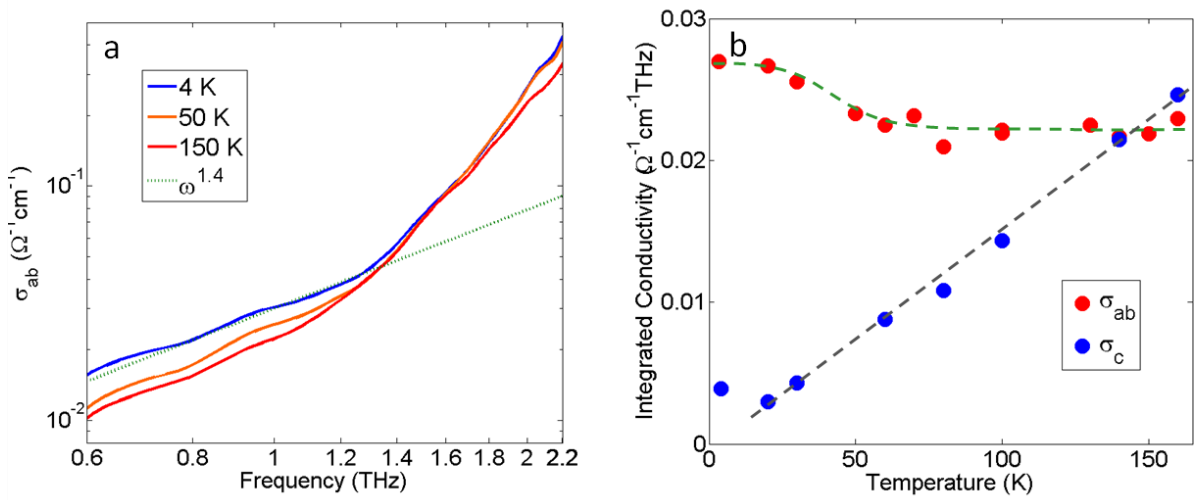

**Figure 2 | Optical conductivity along the kagome (*ab*) planes in Herbertsmithite at various temperatures.** **a.** The in-plane optical conductivity $\sigma_{ab}$ in the spectral range of 0.6 - 2.2 THz at temperatures T = 4, 50 and 150 K. The spectra consist of a higher-frequency component, arising from a phonon resonance at ~3 THz, and a lower-frequency component that exhibits a power-law dependence on frequency as $\sigma_{ab} \sim \omega^\beta$ with $\beta \approx 1.4$ (dotted line). The data are plotted in log-log scale to highlight this power-law behavior. **b.** The integrated values of the in-plane conductivity $\sigma_{ab}$ (Figure 2a) and out-of-plane conductivity $\sigma_c$ (Figure 3) from 0.6 to 1.4 THz at temperatures T = 4 - 160 K. $\sigma_{ab}$ is found to decrease as the temperature increases up to T ~ 60 K, where it remains constant. In contrast, $\sigma_c$ increases monotonically with the temperature. Dashed lines are guides to the eyes.

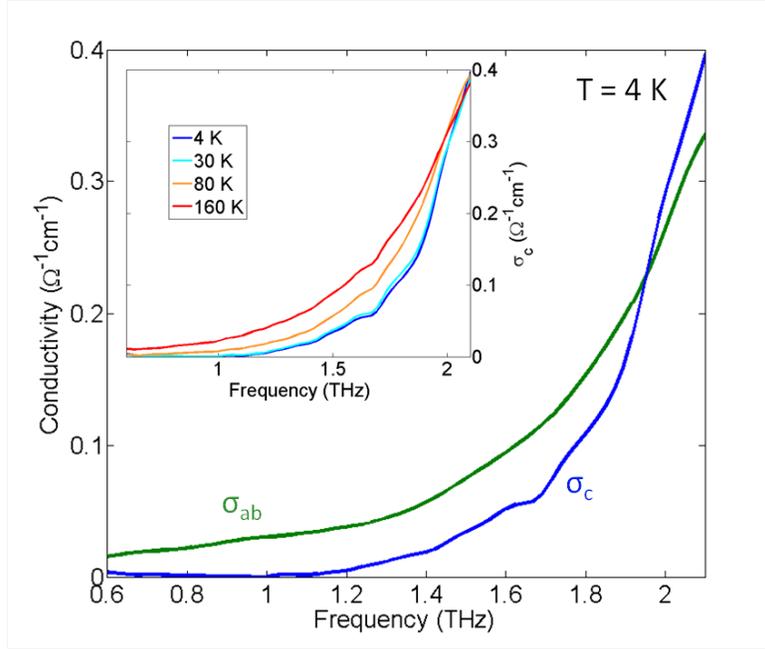

**Figure 3 | Out-of-plane optical conductivity of Herbertsmithite.** The out-of-plane conductivity spectrum $\sigma_c$ (blue line) is plotted in the spectral range of 0.6 - 2.2 THz at T = 4 K, in comparison with the in-plane conductivity $\sigma_{ab}$ (green line). The inset displays the $\sigma_c$ spectra at T = 4, 30, 80 and 160 K. The magnitude of $\sigma_c$ is found to increase with the temperature.

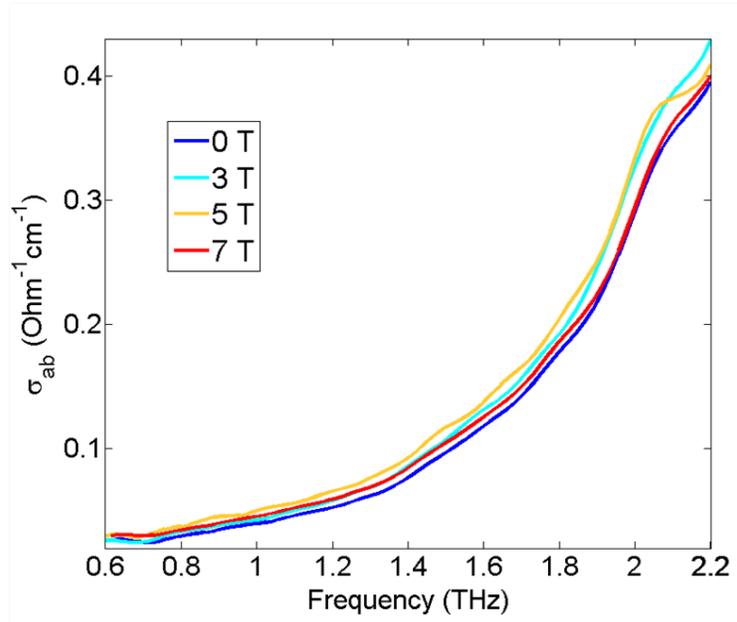

**Figure 4 | In-plane conductivity at various magnetic fields.** The conductivity $\sigma_{ab}$ was measured in the spectral range of 0.6 - 2.2 THz at magnetic fields of 0, 3, 5, and 7 T. No systematic magnetic field dependence is observed. The small variation between measurements is attributed to experimental uncertainties.

# Supplementary Information

## S1. Extraction of in-plane and out-of-plane conductivity

In the measurement of the in-plane ($\sigma_{ab}$) and out-of-plane ($\sigma_c$) conductivity, the polarization of the terahertz (THz) radiation was aligned along the ordinary and the extraordinary optical axis of the sample, respectively. By measuring the transmitted THz field with and without the samples in the beam path, we obtained the complex transmission coefficient ($t$) of the sample. The complex refractive index ($n$) can be readily extracted from the transmission by the following equations[1]

$$t = \frac{4n e^{in\omega d/c}}{(n+1)^2} \quad (1)$$

$$\frac{1}{n^2} = \frac{\cos^2(\theta)}{n_c^2} + \frac{\sin^2(\theta)}{n_{ab}^2} \quad (2)$$

Here $d$, $\omega$, $n_{ab}$ and $n_c$ are, respectively, the sample thickness (0.8 mm), the incident photon frequency, and the in-plane and out-of-plane complex refractive index. $\theta$ denotes the angle between the THz polarization and the $c$-axis of the crystal. The optical conductivity $\sigma$ is related to $n$ as:

$$\sigma = 2\varepsilon_0 \omega \operatorname{Re}(n) \operatorname{Im}(n) \quad (3)$$

In the measurement of in-plane conductivity, $\theta = 90°$ and $n = n_{ab}$. $\sigma_{ab}$ can thus be extracted easily by using equation (2). In the case of out-of-plane conductivity $\sigma_c$, due to the inclination of the $c$-axis with respect to the sample surface (Figure 1b in the main paper), the polarization of the incident THz field makes an angle of $\theta = 23°$ with the $c$-axis of the crystal in our measurement. The transmission thus involves both the in-plane ($n_{ab}$) and out-of-plane ($n_c$) refractive indices. The out-of-plane refractive index ($n_c$) can be calculated using the experimentally measured $n$ and the previously measured in-plane refractive index ($n_{ab}$) using equation (2). We also note that in the measurement of the out-of-plane conductivity, the THz pulse travels inside the sample with a group velocity slightly deviated from its wave vector due to the anisotropy of the crystal. However, we estimate that the deviation angle is rather small (~3°) and the resultant error in the transmission measurement is negligible (~0.3%).

## S2. Measurement of reflectance in the spectral range of 3 - 16 THz

As shown in Figure 2a of the main paper, we observed in the in-plane conductivity spectra $\sigma_{ab}(\omega)$ a higher-frequency component, which is significant at frequencies above 1.4 THz. We ascribe this component to phonon absorption, distinct from the spin-induced absorption that dominates at frequencies below 1.4 THz. In order to elucidate the nature of the higher-frequency absorption component, we have investigated the optical response of the Herbertsmithite crystal at frequencies well above 1.4 THz. In particular, we measured the reflectance of the sample in the spectral range of 3 - 16 THz by means of Fourier transform infrared (FTIR) spectroscopy. The measurements were performed at room temperature and ambient conditions. The polarization of the incident infrared beam was aligned along either the ordinary or extraordinary axis of the crystal for the measurement of the in-plane and out-of-plane reflectance, respectively.

Figure S1a displays the in-plane reflectance of Herbertsmithite at frequencies above 3 THz. The spectrum exhibits four peaks, which we attribute to four phonon absorption resonances. We can describe the experimental data adequately by considering reflection arising from four Lorentzian oscillators with a small constant reflection background (red line in Figure S1a). The lowest-frequency resonance in the fit is at 3.4 THz. From the reflectance data, we have extracted the optical conductivity and extrapolated it to frequencies below 2 THz in order to make comparison with the conductivity data obtained by THz time-domain spectroscopy. As shown in Figure S1b, the extrapolated conductivity agrees reasonably well with the THz data for both the absorption magnitude and spectral shape. Therefore, our analysis strongly suggests that the higher-frequency component observed in the THz in-plane conductivity is a tail of a phonon resonance at ~3 THz.

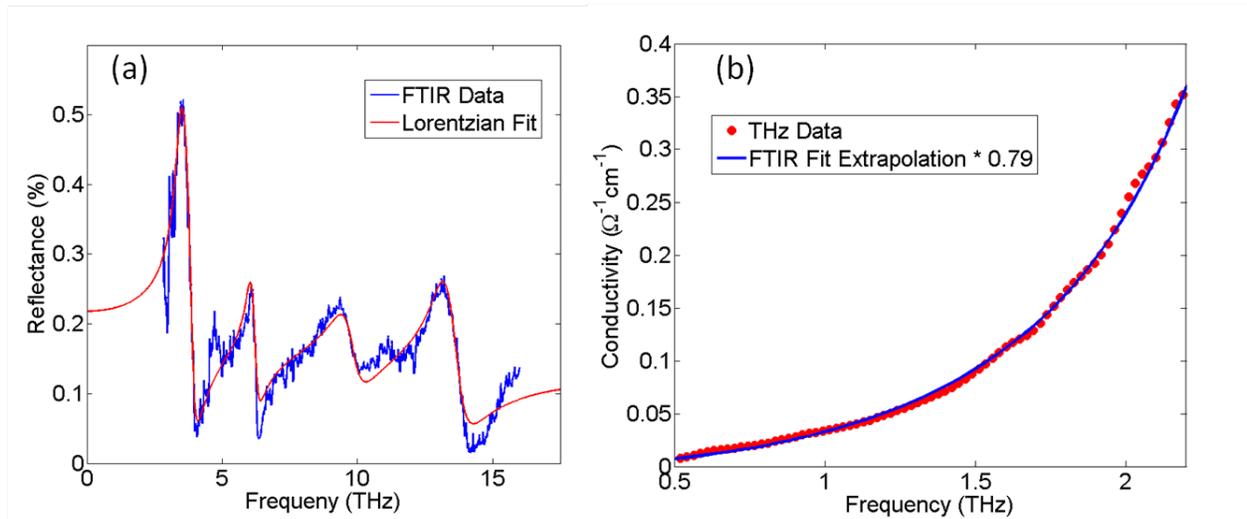

**Figure S1.** (a) In-plane reflectance spectrum measured by FTIR spectroscopy. The data are compared with a fit considering four Lorentzian oscillators (red line). The measurement was carried out in room temperature and ambient conditions. (b) Extrapolation of the extracted conductivity from the Lorentzian fit in (a), in comparison with the in-plane conductivity measured by THz spectroscopy at room temperature. The FTIR extrapolation data was scaled by a factor of 0.79 to reach the best agreement.

### S3. Measurement of optical conductivity under magnetic field

According to previous $x$-ray scattering studies[2], a 5% excess of copper atoms were found between the kagome planes in our Herbertsmithite sample. Although these additional copper atoms leave the geometric frustration intact, they can act as random paramagnetic spins to induce a Curie tail in magnetic susceptibility[3,4]. In respect to our experiment, these spin impurities may contribute to the optical conductivity in the low frequency range, and hence obscure the interpretation of our experimental data. An investigation of the magnetic field dependence should be able to rule out their effect. In the presence of magnetic field, the conductivity due to the excess copper atoms is expected to disappear at low frequency due to the Pauli Exclusion Principle as the defect spins are aligned to the magnetic field. In contrast, the spin liquid state cannot be magnetically aligned except at fields greater than 12 T or at temperatures below 1 K, where a phase transition from spin liquid to spin solid may occur[5,6]. In our experiment, we have measured the in-plane conductivity $\sigma_{ab}(\omega)$ in the spectral range of 0.6 - 2.2 THz under magnetic fields from 0 T to 7 T at T = 6 K (Figure S2). We did not observe any systematic dependence of the absorption on the magnetic field. In particular, the exponent of the power-law fit to the data is essentially unchanged with the magnetic field. The slight change of the absorption magnitude is within our experimental errors, including the possible minor clipping of the incident beam on the sample holder and the imperfect alignment of the optics within the chamber of the superconducting magnet. From this result, we can safely exclude magnetic disorder as a source of the low-frequency optical conductivity in our experiment.

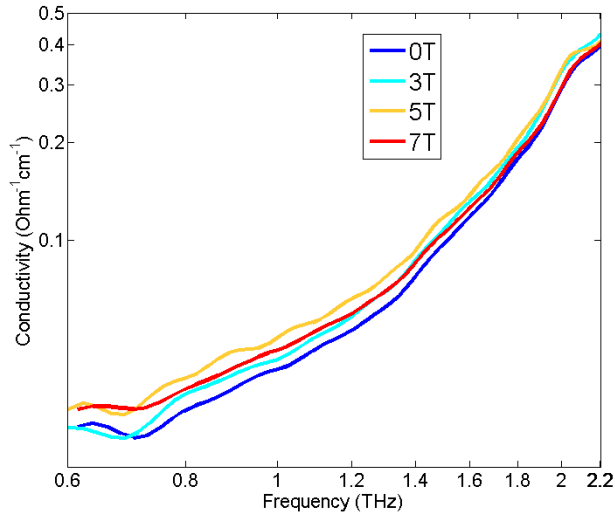

**Figure S2.** In-plane optical conductivity of Herbertsmithite at different magnetic fields. The spectra are reproduced from Figure 4 in the main paper, and here plotted in the log-log scale to highlight the power-law behavior of the conductivity. The measurement was performed at T = 6 K.

**Supplementary references:**